\documentstyle[11pt,paspconf,epsfig]{article}
\def\ltsima{$\; \buildrel < \over \sim \;$}
\def\lsim{\lower.5ex\hbox{\ltsima}}
\def\gtsima{$\; \buildrel > \over \sim \;$}
\def\gsim{\lower.5ex\hbox{\gtsima}}

\begin{document}

\title{Bursts from internal shocks: is it really synchrotron emission?}

\author{Annalisa Celotti}
\affil{S.I.S.S.A., via Beirut 2-4, I-34014 Trieste, Italy}

\author{Gabriele Ghisellini}
\affil{Osservatorio Astron. di Brera, via Bianchi 46, I-23807 Merate,
Italy}

\begin{abstract}
Within the standard internal shock scenario, synchrotron emission
would produce a spectrum with slope $F_\nu \propto \nu^{-1/2}$, as
immediate consequence of the cooling timescale being shorter than the
integration time. This is in disagreement with the harder observed
spectra, indicating that a different mechanism is responsible for the
burst emission. Furthermore in this scenario pair production is
expected when photons produced by inverse Compton emission are taken
into account.
\end{abstract}


\section{Introduction}

As amply discussed during this meeting, within the standard scenario
dissipation of bulk kinetic energy occurs at internal shocks,
generated by e.g. the non--uniform emission from the central source,
and subsequently at the shock formed when the fireball impinges onto
the external medium. These dissipation events are observationally
identified with the burst and the following afterglow radiation,
respectively.

It is also quite generally accepted (see however Thompson 1994; Liang 1997;
Ghisellini \& Celotti 1999 and these proc.; Stern, these proc.) that
the radiation is produced through synchrotron emission, although
quantitative (and rather detailed) comparison of spectral and time
evolution has been possible only for the afterglow phase (e.g. Sari,
these proc.). For the burst itself, the observational support is quite
weak except for the prediction of a typical energy of $\sim$ few
hundred keV, which seems indeed to characterize the GRB peak emission.

In what follows we point out that adopting the parameters of the
internal shock scenario to interpret the GRB emission itself leads to:
a) a clear discrepancy between the predicted and observed spectra; b)
copious pair production. 

\subsection{The standard scenario}

The complex variability patterns typical of GRB are attributed to the
emission from ``shells'' of matter ejected from the central engine,
e.g.  with different Lorentz factors, interacting. The faster shell
would reach the slower one at typically $R_{\rm i}\sim \, R_{\rm
o}\Gamma^2$, forming a shock, where bulk energy would be dissipated
through acceleration of protons, electrons and amplification of
magnetic fields.  In particular, it is assumed that electrons are
energized instantaneously to a typical Lorentz factor which
corresponds to equipartition with the other forms of energy:
$\gamma_{\rm eq}\simeq \epsilon_{\rm e} (m_{\rm p}/m_{\rm e}) (n_{\rm
p}/n_{\rm e})$, where $n_{\rm p}$ and $n_{\rm e}$ are the densities of
proton and electrons and it is assumed that $n_{\rm p} = n_{\rm e}$,
i.e.  there is not a significant amount of electron--positron pairs.
These electrons would then radiate through synchrotron.  Whether the
magnetic field transports a significant fraction of the total power as
Poynting flux or shares a fraction $\epsilon_{\rm B}$ of the energy
which is randomized in the internal shock, its estimated value is $B\,
\simeq \, ( 2\epsilon_{\rm B}/c L_{\rm s})^{1/2}\, \Gamma^{-1}
R^{-1}$, where $L_{\rm s}$ is the synchrotron radiated luminosity.

Thus the typical burst peak frequency is predicted to be 
\begin{equation}
\nu_{\rm s}\, \sim\, 2 \epsilon_{\rm B}^{1/2}\epsilon_{\rm e}^2 L_{\rm
s,48}^{1/2} R_{\rm o,7}^{-1} \Gamma_2^{-2} (1+z)^{-1}\,\, {\rm MeV}
\end{equation}
in good agreement with observations. This constitutes one of the most
robust supports to the standard scenario accounting for the burst
emission. This agreement nevertheless requires that both field and
electron energies are close to equipartition (i.e. $\epsilon_{\rm B}
\sim \epsilon_{\rm e} \sim 1$ \footnote{ $\epsilon_{\rm B}$ and
$\epsilon_{\rm e}$ have been estimated only by applying the
synchrotron scenario to the afterglow/external shock phase.}) and that
$\Gamma$ is constrained within a tight range of values.

\section{The predicted (integrated) spectrum} 

The main point we want to stress here is that the particle cooling
timescales are much shorter than the integration timescale, and as a
direct consequence, the predicted synchrotron spectrum in the entire
X--ray band should have a slope $F_\nu\propto \nu^{-1/2}$, in clear
conflict with observations.

In fact, the radiative timescale of an electron radiating via
synchrotron (and self--Compton) in this scenario is:
\begin{equation}
t_{\rm cool}\, \sim \, 10^{-7} \epsilon_{\rm e}^3 \Gamma_2 \nu_{\rm
MeV}^{-2} (1+U_{\rm r}/U_{\rm B})^{-1} (1+z)^{-1}\, \, {\rm s}
\end{equation}
which is always much smaller than the typical integration time (of the
order of 1 s). Thus one always observe the emission by cooled
particles. As the particle distribution at each time
$N(\gamma,t)\propto \gamma^{-1}$ (to conserve the particle number),
when integrated (i.e. weighted over the cooling timescale) gives
$N(\gamma)\propto\gamma^{-2}$. The corresponding (observed) spectrum
would then have a slope $F_{\nu}\propto \nu^{-1/2}$.

\subsection{Possible alternatives?}

The above result seems inevitable within the specific assumptions of
the equipartition synchrotron scenario. Let us consider alternative
hypothesis which would allow to avoid the above conclusion. The
simplest possibility is to envisage a situation where energy
equipartition is not reached and $\epsilon_{\rm B}$ and/or
$\epsilon_{\rm e}$ are $\ll$ 1.  However, the following considerations
apply:

\begin{itemize}

\item in either cases (i.e. $\epsilon_{\rm B}$ or $\epsilon_{\rm e}\ll 1$) 
both parameters have to be orders of magnitudes smaller than
the equipartition value in order for $t_{\rm cool}$ to be comparable
to the integration time. No typical peak energy for the burst would
then be expected;

\item if $\epsilon_{\rm B}\ll 1$, although the synchrotron cooling
would be slower, the amount of radiation energy density in the
emitting region - which necessarily corresponds to the observed fluxes
($U_{\rm rad} \sim L_{\rm s}/(4\pi c R^2 \Gamma^2))$ - implies that
the inverse Compton cooling timescale would be much shorter than the
integration time, leading again to a steep spectrum;

\item a similar effect (i.e. increase of synchrotron cooling timescale
but decrease in the inverse Compton one) would occur if the (strong)
equipartition field was limited to a thin spatial region, so that
electrons would not loose more than a small fraction of their energy
before escaping it;

\item $\epsilon_{\rm e} \ll 1$ would lead to a very inefficient
radiative dissipation, as the energy dissipated in the shocks is
already assumed to accelerate all of the available particles;

\item electrons could be continuously re--heated, thus avoiding the
formation of a cooled particle distribution.  As it is not possible to
re--accelerate the very same particles (as this would exceed the total
energetics), one has to assume that only `selected' electrons are
continuously accelerated for the entire duration of the shell--shell
interaction.  A strong fine tuning is then required, as both their
number ($\sim$ total number of particles times the cooling time), and
their energy ($\sim \gamma_{\rm eq}$ although equipartition would not
be reached) would be determined;

\item even for a power law distribution of particles $\propto
\gamma^{-p}$, resulting from continuous heating and cooling, a
spectrum steeper than $F_{\nu}\propto \nu^{-1/2}$ is implied, as only
a very small fraction of particles can be accelerated to high
energies.

\end{itemize}

We conclude that the time integrated spectrum predicted by the
standard scenario is steeper than what observed.

\section{Pair production}

The second effect which significantly alters the spectral predictions
of the standard scenario is pair production. In fact, within the
equipartition hypothesis, the Compton parameter of the emitting zone,
whose thickness is determined by the electron cooling length, is of
the order of unity.  This implies that an amount of luminosity
comparable with the synchrotron one is dissipated through inverse
Compton scattering.  The Compton component is going to be peaked at a
(comoving) energy $ \nu_{\rm c}\, \sim \, 90 \epsilon_{\rm
B}^{1/2}\epsilon_{\rm e}^4 L_{\rm s,48}^{1/2} R_{\rm o,7}^{-1}
\Gamma_2^{-3} \,\, {\rm GeV}$. It is therefore crucial to estimate the
possible role of photon--photon interactions leading to
electron--positron pair production.  The optical depth for this
process is proportional to the compactness in target photons:
\begin{equation}
\ell\,\simeq \, 270 {L_{48}\over \Gamma_2^2 R_{13}}\, {\Delta R \over R}
\end{equation}
where $\Delta R$ is the travel path and $1/\Gamma$ is the typical
angle between the interacting photons.  Then $\tau_{\gamma\gamma} \sim
\ell/60$ for observed photon energies $h\nu\sim \Gamma m_ec^2$ (and
depends on frequency as $\nu^\alpha$ for larger energies).

If we consider the region in front of the emitting shell $\Delta R
\sim R$ and hence $\ell$ is large: all photons above threshold would
be absorbed and produce pairs, which in turn radiate and give raise to
a cascade.  The qualitative results (for a stationary source) are
that: a) $\gamma$--rays are reprocessed into lower energy photons,
leading to a steeper spectrum; b) leptons are created in large number,
implying that the average energy per particle has to decrease.

\section{Conclusions}

Within the frame of the internal shock scenario synchrotron (and
inverse Compton) emission do not seem to reproduce the spectrum of the
burst itself. This directly follows from the requirement of short
cooling timescales intrinsic to this scenario. Furthermore contrary to
its assumptions, pair production is expected to occur.  Even the
relaxation of the equipartition hypothesis does not simply allow to
overcome the difficulties of the model. (For a more detailed analysis
see Ghisellini, Celotti \& Lazzati 1999, in prep.)

Alternative hypothesis on the dissipation/particle acceleration and/or
radiation processes seem to be required.  Comptonization by a
quasi--thermal particle distribution appears to be a promising
possibility (Thompson 1994; Ghisellini \& Celotti 1999 and these
proc.; Stern, these proc.).

\section*{Acknowledgments} It is a pleasure to thank to the organizers
for a very stimulating meeting and a fantastic `environment'.  AC
acknowledges the Italian MURST for financial support and ITP, Santa
Barbara, for hospitality during the writing of this contribution. This
research was supported in part by the National Science Foundation under
Grant No. PHY94-07194.

\end{document}